\documentclass[11pt]{article} 
\pdfoutput=1
\usepackage{amsmath}
\usepackage{amssymb}
\usepackage{setspace}

\usepackage{geometry} 
\usepackage{graphicx} 

\usepackage{booktabs} 
\usepackage{array} 
\usepackage{paralist} 
\usepackage{verbatim} 
\usepackage{subfig} 

\usepackage[nottoc,notlof,notlot]{tocbibind} 

\title{Constraints on Axion Inflation\\
from the Weak Gravity Conjecture}
\author{Tom Rudelius\\ \\
  \small Jefferson Physical Laboratory, Harvard University,\\
\small  Cambridge, MA 02138, USA\\
  }

\begin{document}
\setlength{\baselineskip}{16pt}
\begin{titlepage}
\maketitle
\begin{spacing}{1.5}
\vspace{-36pt}
\begin{abstract}
We derive constraints facing models of axion inflation based on decay constant alignment from a string-theoretic and quantum gravitational perspective.  In particular, we investigate the prospects for alignment and `anti-alignment' of $C_4$ axion decay constants in type IIB string theory, deriving a strict no-go result in the latter case.  We discuss the relationship of axion decay constants to the weak gravity conjecture and demonstrate agreement between our string-theoretic constraints and those coming from the `generalized' weak gravity conjecture.  Finally, we consider a particular model of decay constant alignment in which the potential of $C_4$ axions in type IIB compactifications on a Calabi-Yau three-fold is dominated by contributions from $D7$-branes, pointing out that this model evades some of the challenges derived earlier in our paper but is highly constrained by other geometric considerations.
\end{abstract}
\end{spacing}
\thispagestyle{empty}
\setcounter{page}{0}
\end{titlepage}

\maketitle

\section{Introduction}

Inflation \cite{guth, linde, albrecht&steinhardt} has become the dominant paradigm for resolving the problems facing early universe cosmology.  Recent measurements of B-mode polarization by the BICEP2 collaboration \cite{bicep2} have led to increased interest in the models of large-field inflation (see e.g. \cite{axion1, axion2, axion3, axion4, axion5, axion6, McAllisterLong, axion8, Pramod, axion9, axion10, axion11, axion12, axion13, Sundrum}).  The majority of the signal has since been explained by dust measured by the \emph{Planck} collaboration \cite{Planckdust, PlanckBicep}, but if inflation in our universe did proceed by a large-field mechanism like chaotic or natural inflation, a measurement of primordial B-modes is likely to take place in the very near future \cite{Zaldarriaga}.

Large-field inflation presents problems from the effective field theory perspective because one expects Planck-suppressed operators to generate large corrections to the potential, rendering inflation unnatural \cite{Lotfi}.  By far the most popular solution to this problem is to use axions, whose shift symmetry protects their potential from perturbative corrections.  Nonetheless, this shift symmetry is broken by non-perturbative effects which are sensitive to ultraviolet physics.  In \cite{Banks}, it was shown that various attempts in string theory to produce an axion with decay constant large enough for inflation fail.  In the years since, theorists' attempts to evade this issue have centered around three proposals:
\begin{itemize}
\item $N$-flation \cite{Nflation}
\item Axion monodromy inflation \cite{AxionMonodromy}
\item Axion decay constant alignment \cite{KNP}
\end{itemize}
Challenges to the first of these proposals were discussed in \cite{Rudelius}.  In the present work, we present similar challenges to the third, finding constraints that are unlikely to be satisfied in typical Calabi-Yau compactifications of string theory but may well be avoided in certain cases.  The paper is structured as follows: in \S 2, we introduce the proposal and offer a critique of the philosophy which has motivated it.  In \S 3, we discuss the possibility of axion decay constant alignment in type IIB compactifications on Calabi-Yau three-folds.  In \S 4, we introduce the weak gravity conjecture and highlight its implications for decay constant alignment.  In \S 5, we study a notable attempt to realize decay constant alignment in type IIB compactifications \cite{McAllisterLong} and point out that it actually avoids some of the no-go results derived earlier in the paper, but it faces other geometric constraints which one rarely (if ever) expects to be satisfied in a Calabi-Yau compactification.  Finally in \S 6, we make some final remarks and present some interesting directions for future research.

\section{A Critique of Alignment Philosophy}

In \cite{KNP}, Kim, Nilles, and Peloso developed a mechanism which purported to show that one could achieve an effective large axion scale from an originally small scale.  We point out that this conclusion is the result of a particular basis choice, which conceals the fact that a large scale is actually required.  One can always move to a basis in which this scale is manifest, and it seems unlikely that string theory, in its distaste for large axion decay constants \cite{Banks, Rudelius}, will be fooled by moving into the original basis.

Consider two axions, $\phi_1$ and $\phi_2$, with shift symmetries broken by the potential,
\begin{equation}
V = \Lambda_1^4 \left( 1 - \cos{(\frac{\phi_1}{f_1} + \frac{\phi_2}{f_2})}\right) + \Lambda_2^4 \left( 1 - \cos{(\frac{\phi_1}{g_1} + \frac{\phi_2}{g_2})}\right).
\end{equation}
Here, the $\Lambda_i$ are dynamically-generated scales, and the $f_i, g_i$ are decay constants.  The standard lore, based on the observation of \cite{Banks}, is that these decay constants are required to be sub-Planckian, $f_i, g_i \lesssim M_p$.  The Kim-Nilles-Peloso (KNP) mechanism is based on the following observation: if $f_1 / f_2 = g_1 / g_2$, then one will get a flat direction in the potential.  Of course, this would correspond to an unbroken shift symmetry, which is not allowed in a theory of quantum gravity and so must be broken.  However, if we take the equality to be approximate, $f_1 / f_2 \approx g_1 / g_2$, then the shift symmetry will be broken and a long, flat direction in axion moduli space will emerge.  Even if $f_i, g_i \sim M_p$, this flat direction can still be much longer than $M_p$ if the alignment is sufficiently good.

The claim, then, is that one can achieve a super-Planckian displacement from sub-Planckian decay constants.  However, this is an ambiguous statement--as we will now see, it is highly sensitive to the choice of basis, which should have no effect on the physics.  Without loss of generality, take the decay constants $g_i$ to be given by $g_1 : = g$, $g_2= g (\frac{f_2}{f_1} - \delta)$, $f_1 \neq 0$.  When $\delta \rightarrow 0$, the KNP mechanism comes into effect and produces a long flat direction in axion moduli space, with perfect alignment given by $\delta = 0$.  Now, define a new basis by making a clockwise rotation by an angle $\theta = \tan^{-1}(f_1/f_2)$ in the $\phi_1, \phi_2$ plane,
\begin{equation}
\tilde{\phi}_1 = \cos(\theta)\phi_1 + \sin(\theta)\phi_2\,,~~~~\tilde{\phi}_1 = -\sin(\theta)\phi_1 + \cos(\theta)\phi_2.
\end{equation}
As a result, the potential becomes,
\begin{equation}
V = \Lambda_1^4 \left( 1 - \cos{(  \frac{f}{f_1 f_2} \tilde\phi_1)} \right) + \Lambda_2^4 \left( 1 - \cos{(\frac{f + \mathcal{O}(\delta)}{f_2 g} \tilde\phi_1+ \frac{f_1^2 \delta + \mathcal{O}(\delta^2)}{f f_2 g} \tilde\phi_2)}\right),
\end{equation}
where $f := \sqrt{f_1^2 + f_2^2}$.  Examining the axion decay constants in this new basis, we see that whenever $f_i, g \sim M_p$, the decay constant $\tilde{g}_2 = \frac{f f_2 g}{f_1^2 \delta}$ blows up as $\delta \rightarrow 0$.  There is a scale of order $\frac{1}{\delta} M_p$ hidden by the basis choice $\phi_1, \phi_2$ that becomes manifest upon switiching to the $\tilde\phi_1, \tilde\phi_2$ basis.

What are we to make of this?  The axion `scale' is a basis-dependent quantity and is therefore unphysical.  A bound on a physical theory should presumably involve physically meaningful quantities.  The most obvious physical quantities lying around are the volume and radius of the moduli space.  Assuming the apparent bound $V_1=r_1 \lesssim M_p$ in the single axion case, the most natural bound on the volume is (a) $V_N \lesssim M_p^N$, and the most natural bounds on the radius are either (b) $r_N \lesssim \sqrt{N} M_p$ (the na\"ive $N$-flation expectation) or (c) $r_N \lesssim M_p$, where $N$ is the number of axions.  In previous studies, (a) has been assumed (often implicitly) to be the correct bound; a bound on the moduli space volume would not induce any bound on the radius, and one could attain an arbitrarily large inflaton traversal given sufficient alignment.  However, in \cite{Rudelius}, evidence began to mount that the option (c) is the right one.  If correct, this would put natural inflation in tension with quantum gravity, for the usual $N$-flation scenarios and decay constant alignment scenarios both involve an axion moduli space whose radius is much larger than $M_p$.

However, there is a possible way to get around this argument and realize inflation with axions.  Even if the radius of axion moduli space is bounded, one could in principle introduce additional contributions to the potential which dominate over the ones defining the boundary of moduli space.  If these contributions are sufficiently smooth, they may give rise to prolonged inflation in which an axion traverses its moduli space several times.  The most famous example along these lines is the original axion monodromy scenario, but we will see that there may be additional examples which give rise to natural inflation rather than chaotic monomial inflation.

\section{Decay Constant Alignment in Type IIB String Theory \label{sec:STRING}}

In this section, we address the possibility of decay constant alignment in type IIB compactifications on a Calabi-Yau three-fold.  Such a compactification yields an $\mathcal{N}=2$ supergravity theory in 4d.  The R-R axion $C_0$ descends to an axion in $4d$, and pairs up with the dilaton $\Phi$ to form the complex axiodilaton,
\begin{equation}
\tau = C_0 + i e^{-\Phi}.
\end{equation}
Additional axions arise from integrating the NS-NS 2-form $B_2$, the R-R 2-form $C_2$, and the R-R 4-form $C_4$ over cycles of the appropriate dimensionality,
\begin{equation}
b_i = \frac{1}{2 \pi\alpha'}\int_{\Sigma_{i}}{B_2}\,,~~~~c_i = \frac{1}{2 \pi\alpha'}\int_{\Sigma_{i}}{C_2}\,,\vartheta_i = \frac{1}{2 \pi (\alpha')^2}\int_{\Sigma_{i}}{C_4}.
\end{equation}
The axions' shift symmetries are broken due to instanton effects.  K\"{a}hler moduli similarly arise from integrating the K\"{a}hler form $J$ over the same 2-cycles,
\begin{equation}
t_i = \frac{1}{\alpha'}\int_{\Sigma_{i}}{J}.
\end{equation}
A study of the $b_i$ and $c_i$ axions was carried out in \cite{Rudelius}, where it was argued that axion moduli space radii much larger than $M_p$ are forbidden.  However, the $\vartheta_i$ axions are more difficult to deal with, and although \cite{Rudelius} showed that the scaling of the metric on $\vartheta_i$ axion moduli space does not lend itself to the possibility of $N$-flation, it did not rule out the possibility of decay constant alignment.  Furthermore, it was precisely with regard to the $\vartheta_i$ axions that the moduli stablization scenario of \cite{CicoliDutta} pertained, and so a more detailed analysis here is not merely an academic exercise.

We begin our study of decay constant alignment by considering compactification on a Calabi-Yau three-fold $Z$ with $h^{1,1}=2$.  To get a realistic model of nature with chiral matter and $\mathcal{N}=1$ supersymmetry, one would have to insert orientifold planes under whose orientifold action the cohomology group $H^{1,1}$ splits into a parity-even and parity-odd part \cite{BaumannMcAllisterBook},
\begin{equation}
H^{1,1}(Z) = H_+^{1,1}(Z) \oplus H_-^{1,1}(Z).
\end{equation}
The $\vartheta_i$ axions correspondingly split under this $\mathbb{Z}_2$ action according to whether or not their associated $4$-cycles are even or odd.  In a theory with $O3/O7$ orientifold planes, the $\vartheta_i^+$ axions (i.e. those associated with even cycles) are projected in, whereas the $\vartheta_i^-$ axions are projected out.  We henceforth take $h^{1,1}=h^{1,1}_+=2$, $h^{1,1}_-=0$, so our analysis applies to both the $\mathcal{N}=2$ and $\mathcal{N}=1$ scenarios.

In these compactifications, Euclidean $D3$-branes wrapping effective 4-cycles give rise to potential contibutions for the $\vartheta_i$ axions of the form,
\begin{equation}
V \supset \sum_{n} \mathcal{A} e^{- n \tau_i/g_s} \left( 1 - \cos{n \vartheta_i / g_s}\right).
\label{Veq}
\end{equation}
$\mathcal{A}$ is usually estimated to be of order $M_p^4$.  We are here neglecting instanton corrections to the metric.  This assumption is justified as long as we are in the limit in which all of the irreducible 2-cycles and 4-cycles have volume $\gtrsim 1$ in string units, and although in principle these unknown corrections could lead to an increased moduli space radius, this proposal is independent of the decay constant alignment mechanism, and so we do not concern ourselves with it here.  Furthermore, unless $g_s$ (and hence the axion decay constant) is very small, the exponential of (\ref{Veq}) is not suppressed in the limit that the cycle volumes $\tau_i$ are taken small, so maintaining an inflaton sufficiently light for inflation while also maintaining a sufficiently large decay constant is difficult in the $\tau_i \rightarrow 0$ limit.  There are also multi-instanton terms in the potential coming from $D3$-branes wrapping some number of times around two different cycles, but in the limit $\tau_i \gtrsim 1$ these will produce subdominant contributions to the potential which can only shrink the field range accessible to the inflaton.  Since we are concerned with an upper bound on the length of an inflaton traversal, we neglect these terms.

\begin{figure}
\begin{center}
\includegraphics[trim=10mm 5mm 10mm 10mm, clip, width=70mm]{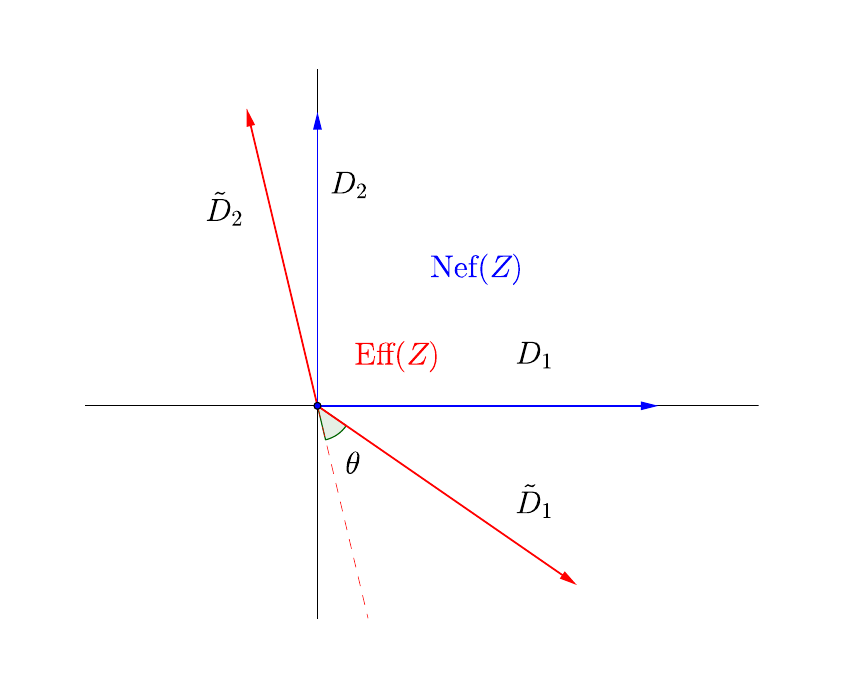}
\caption{The effective cone (red) is spanned by two irreducible effective divisors.  The nef cone (blue) is contained in the closure of the effective cone.  `Anti-alignment' occurs if the angle $\theta$ between $\tilde{D}_1$ and $\tilde{D}_2$ approaches $\pi$.}
\label{figeffective}
\end{center}
\end{figure}

There are two cones in the vector space of divisors $N^1(Z)$ that are relevant for our discussion: the nef cone and the effective cone of divisors.  The nef cone Nef($Z$) is the cone of nef divisors i.e. divisors $D$ satisfying $D \cdot C \geq 0$ for all irreducible effective curves $C$ in $Z$.  The effective cone Eff($Z$) is the the cone consisting of linear combinations $\sum_i a_i D_i$ of irreducible effective divisors $D_i$ with $a_i \geq 0$.  The K\"ahler cone is the interior of the nef cone, so the K\"ahler form is given by $J=\sum_i t^i D_i$ where $D_i$ generate the nef cone and $t^i > 0$.  The closure of the effective cone contains the nef cone.  We set $D_1$, $D_2$ to be the generators of the nef cone and $t^1, t^2$ normalized to correspond to the volumes of their dual curves in string units.  We set $\tilde{D}_1 = \alpha D_1 + \beta D_2$ and $\tilde{D}_2 = \gamma D_1 + \delta D_2$ to be the generators of the effective cone, normalized so that $\tilde{D}_i$ is the homology class of the irreducible effective divisor corresponding to this generator, and chosen so that $\alpha > 0$, $\delta > 0$ (see Figure \ref{figeffective}).

We will work in the basis generated by the nef divisors $\{ D_1, D_2\}$, taking these two divisors to be represented by orthonormal vectors in the plane.  In what follows, we will sometimes abuse notation and use the term divisor to describe a divisor's representation in this basis, so that the `length' of a divisor or the `angle between two divisors' refer respectively to the length of the vector representing the divisor and the angle between the two vectors representing the two divisors in the chosen basis.  In this basis, the angle between the axion decay constant vectors is given precisely by the angle $\theta$ between the effective divisors wrapped by the Euclidean $D3$-branes.  Two possibilities for decay constant alignment emerge--$\theta \approx  0$ or $\theta \approx \pi$.  We shall refer to these cases as `alignment' and `anti-alignment,' respectively.  One might worry that we have not included the effects of kinetic mixing.  However,  we may take these into account by moving to a new basis in which the kinetic terms are canonically normalized.  This is a linear transformation on the fields.  Since a linear transformation maps $0^\circ$ angles to $0^\circ$ angles and $180^\circ$ angles to $180^\circ$ angles, the conditions for alignment and anti-alignment are preserved upon moving to a basis with canonically normalized kinetic terms. 

We consider `anti-alignment' first.  Anti-alignment can actually be expected to occur reasonably often.  The effective divisors in the interior of the effective cone can be written as positive linear combinations of the generating divisors $\tilde{D}_1$ and $\tilde{D}_2$.  This means that their volumes will be posititive $\mathcal{O}(1)$ linear combinations of the volumes $V(\tilde{D}_1)$ and $V(\tilde{D}_2)$ and so will often be subdominant.  This means that if the angle between $\tilde{D}_1$ and $\tilde{D}_2$ is approximately $\pi$, or equivalently if $\beta/\alpha \approx \gamma/\delta$, then anti-alignment is likely to occur.

However, we now show that anti-alignment in this situation cannot sustain inflation.  The leading contributions to the potential come from Euclidean $D3$-branes wrapping effective 4-cyles in the geometry.  Which effective 4-cycles contribute, precisely, is determined by homological data of the cycle and flux data of the compactification \cite{Halverson}.  However, the prospects of anti-alignment are enhanced when the angle $\theta$ between the divisors approaches $\pi$.  From Figure \ref{figeffective}, we see that the greatest possible anti-alignment between any two divisors occurs when the two divisors are chosen to be the generators of the effective cone.  In other words the angle between two effective divisors is maximized for the pair $\tilde{D}_1,\tilde{D}_2$.  Thus, to derive an upper bound on the field range that can be attained by anti-alignment, it suffices to consider the potential that arises from Euclidean $D3$-branes wrapping these two cycles,
\begin{equation}
V \supset e^{- V(\tilde{D}_1) / g_s} \left(1 - \cos(\frac{\alpha \vartheta_1 + \beta \vartheta_2}{g_s})\right) + e^{- V(\tilde{D}_2)/g_s} \left(1 - \cos(\frac{\gamma \vartheta_1 + \delta \vartheta_2}{g_s})\right).
\label{eqn9}
\end{equation}
We see that the decay constants are indeed aligned when $\beta/\alpha \approx \gamma/\delta$, as stated.  The volumes are given by
$$
V(\tilde{D}_1) = \tilde{D}_1 \cdot J \cdot J = \displaystyle\sum_{i,j=1}^2{\kappa_{ij1} \alpha t^i t^j + \kappa_{ij2} \beta t^i t^j},
$$
\begin{equation}
V(\tilde{D}_2) = \tilde{D}_2 \cdot J \cdot J = \displaystyle\sum_{i,j=1}^2{\kappa_{ij1} \gamma t^i t^j + \kappa_{ij2} \delta t^i t^j},
\label{volumes}
\end{equation}
where $\kappa_{ijk} = D_i \cdot D_j \cdot D_j$ are the triple intersection numbers in the $\{ D_1, D_2 \}$ basis.  When the volumes are of order $1/N$, harmonics up to order $N$ may become important and cut down the effective field range available to the axion by a factor of $N$.  Furthermore, corrections to the metric from branes wrapping various cycles will become non-negligible, and the mass of the axion will no longer be suppressed relative to the Planck scale, making it phenomenologically unacceptable for inflation.  The fundamental domain of moduli space is the region specified by requiring the arguments of the cosines in (\ref{eqn9}) to be between $-\pi$ and $+\pi$.  If we enforce the large volume condition $V(\tilde{D}_i) \gtrsim 1$ and set $g_s < 1$ to maintain theoretical control of the string loop expansion we may bound this region as,
\begin{equation}
|\alpha \vartheta_1 + \beta \vartheta_2| < \pi  \lesssim \pi V(\tilde{D}_1)\,,~~~~ |\gamma \vartheta_1 + \delta \vartheta_2|< \pi  \lesssim \pi V(\tilde{D}_2).
\label{fieldrange}
\end{equation}
Here, the $+$ sign corresponds to traveling along the diagonal, so that sgn$(\vartheta_1)=$ sgn$(\vartheta_2)$, while the $-$ corresponds to traveling along the anti-diagonal, sgn$(\vartheta_1)=$ $-$sgn$(\vartheta_2)$.  In the case of anti-alignment, the diagonal direction is stretched large, while in alignment, the anti-diagonal direction is stretched out.  For the present case of anti-alignment, we therefore take the $+$ sign in (\ref{fieldrange}).

\begin{figure}
\begin{center}
\includegraphics[trim=9mm 5mm 12mm 5mm, clip, width=60mm]{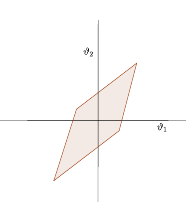}
\includegraphics[trim=9mm 5mm 12mm 5mm, clip, width=60mm]{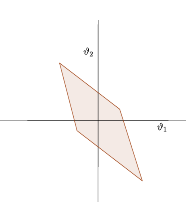}
\caption{When two irreducible divisors are anti-aligned at an angle just less than $\pi$, the field range is stretched so that the greatest displacement occurs when $\vartheta_1$ and $\vartheta_2$ have the same sign (left).  When the irreducible divisors are aligned an an angle of about $\pi$, the greatest displacement occurs when the two axions have opposite sign (right).}
\label{antialignment}
\end{center}
\end{figure}

The metric on field space is given by \cite{Grimm},
\begin{equation}
g_{ij} = \frac{9}{2}\frac{t^i t^j}{(\sum_{k,l,m}\kappa_{klm} t^k t^l t^m)^2} - 3 \frac{(\sum_n\kappa_{ijn}t^n)^{-1}}{\sum_{k,l,m}\kappa_{klm} t^k t^l t^m}.
\label{gij}
\end{equation}
Since the divisors $D_1$ and $D_2$ are nef, the triple intersection numbers $\kappa_{ijk}$ in our basis are necessarily non-negative, which means that no cancellations can occur in the denominator of the metric and yield parametric enhancement of the moduli space radius--any enhancement must be a result of decay constant alignment.  When anti-alignment occurs between the cycles $\tilde{D}_1$ and $\tilde{D}_2$, the field range accessible to the axions is stretched as shown in Figure \ref{antialignment} (left) so that the greatest inflaton displacement occurs when the inequalities in (\ref{fieldrange}) are saturated and both of the quantities in absolute value are positive.  Letting $(\vartheta_1^+, \vartheta_2^+)$ be the values of $(\vartheta_1, \vartheta_2)$ at that point of maximum dispacement, the length in Planck units of the largest inflaton displacement is given by,
\begin{equation}
r = \left(\sum_{i,j} \vartheta_i^+ g_{ij} \vartheta_j^+ \right)^{1/2}.
\end{equation}
However, plugging in the expressions (\ref{fieldrange}), (\ref{gij}) for $\vartheta_i^+$, $g_{ij}$ yields rather simply,
\begin{equation}
r = \pi \sqrt{\frac{3}{2}} M_p.
\label{bound}
\end{equation}
This shows that decay constant anti-alignment is impossible--as the angle between the decay constant vectors approaches $\pi$, the metric eigenvalues and/or cycle volumes must shrink in such a way as to exactly cancel the effects of alignment, so that the inflaton traversal is universally bounded.  In short, decay constant anti-alignment cannot produce sustained inflation.

The arguments against anti-alignment break down in the case of decay constant alignment.  Decay constant alignment occurs whenever the homology classes of two effective divisors are aligned in $N^1(Z)$.  Here, the largest inflaton displacement occurs when $\vartheta_1$ and $\vartheta_2$ have opposite signs, as shown in Figure \ref{antialignment} (right).  However, there are difficulties involved in achieving alignment.  One cannot use multiple windings of the irreducible divisors $\tilde{D}_1$ and $\tilde{D}_2$, for these effects are subdominant (or at most comparable) to those coming from single-wrapping states.  Furthermore, $\tilde{D}_1$ and $\tilde{D}_2$ cannot be aligned in the `good basis' in which the basis vectors generate the nef cone and hence have positive intersection numbers.  This is because the angle between the generators of Nef($Z$) is $\pi/2$ and the nef cone is contained in the closure of the effective cone generated by $\tilde{D}_1$ and $\tilde{D}_2$.  Hence, the angle between these is at least $\pi/2$.

The only remaining possibility for achieving decay constant alignment is if the effective cone contains additional divisors in its interior which introduce contributions to the potential.  This is a very common occurence in the toric-projective Calabi-Yau manifolds famously classified in \cite{KreuzerSkarke}.  However, intersection theory imposes other constraints which limit the extent to which decay constant alignment can occur.  In order for decay constant alignment to occur, the angle $\theta$ between two divisors in the usual $\{D_1, D_2\}$ basis needs to approach 0.  Furthermore, these two divisors must have smaller volumes than any of the other effective divisors that contribute to the potential.  Otherwise, their potential contributions would be subdominant, and the alignment of divisors would not have a significant effect on the potential.

However, we do not expect that the \emph{only} effective divisors yielding potential contributions will typically be aligned at an angle $\theta \ll 1$.  More often, we expect to find a handful of divisors contributing to the potential, some of which are aligned and some of which are not.  For the sake of illustration, suppose the generators of the effective cone $\tilde{D}_1$ and $\tilde{D}_2$ contribute to the potential, along with some third divisor $\tilde{D}_{int}$, which is aligned with $\tilde{D}_1$.  The volume of $\tilde{D}_{int}$ can be expressed as a positive linear combination of the volumes $V(\tilde{D}_1)$ and $V(\tilde{D}_2)$, $V(\tilde{D}_{int}) = a V(\tilde{D}_1)+b V(\tilde{D}_2)$, $a,b>0$.  For $\tilde{D}_1$ and $\tilde{D}_{int}$ to yield an alignment potential, the volumes of these divisors must be larger than that of $\tilde{D}_2$, hence $a$ and $b$ must be small.  But as $a$ and $b$ shrink, there is another problem: the triple intersection numbers between $\tilde{D}_{int}$ and the other effective divisors must be integers provided the Calabi-Yau is smooth.  In type IIB string theory, singular phases are suppressed by $\alpha'$ \cite{WittenMF, MorrisoncitedinWitten}, so that in the large volume limit with $\alpha' \rightarrow 0$, these phases disappear and we are left to consider only smooth Calabi-Yau's with integral intersection numbers.  If we allow the coefficients $a$ and $b$ to grow arbitrarily large, we can take $\theta$ as small as we would like and still be able to ensure integrality of the intersection numbers, but then the potential contributions from $D3$-branes wrapping $\tilde{D}_{int}$ would be negligible.  If we keep $a$ and $b$ small, then there is a limit on how small $\theta$ can be while maintaining integrality of these intersection numbers.   In particular, for $\theta \rightarrow 0$, we do not expect $\tilde{D}_{int}$ to have integral intersection numbers with $\tilde{D}_1$ and $\tilde{D}_2$, as would be required to implement inflation by decay constant alignment.

The story here regarding decay constant alignment is reminiscent of that observed in \cite{Banks} regarding the single axion case and \cite{Rudelius} regarding the many axion one.  A survey of the landscape of axions in string theory suggests that whenever decay constants in string theory start growing larger than $M_p$, higher harmonics become important and cut off the effective field range.  Studies of axion moduli spaces with $h^{1,1} \sim 10$ indicate that whenever the radius of axion moduli spaces starts growing much larger than $M_p$, contributions to the potential from additional cycles are likely to become important and cut off the effective field range once again.  Here, we have observed hints of a similar phenomenon: decay constant anti-alignment from Euclidean $D3$-branes in Calabi-Yau's with $h^{1,1}=2$ has been ruled out.  Decay constant alignment could produce a larger field range, but as the alignment angle is taken smaller and smaller, contributions to the potential from Euclidean $D3$-branes wrapping other divisors (e.g. $\tilde{D}_2$, in the example of the previous paragraph) are likely to become important and spoil the attempted enhancement.  It may well be possible to achieve sufficient alignment to attain the inflaton traversal of $\mathcal{O}(15 \, M_p)$ needed for agreement with the measured values of $n_s$ in some compactification, but the necessary constraints do not appear likely to be satisfied in a typical compactification.  This suggests that deeper quantum-gravitational constraints may be at play here.

\begin{figure}
\begin{center}
\includegraphics[trim=8mm 8mm 8mm 8mm, clip, width=100mm]{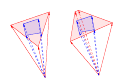}
\caption{When the generators of the effective cone (shown in red) are not aligned, it is possible for the effective cone to be simplicial and still contain the nef cone (shown in blue) (left).  However, if two generators of the effective cone are approximately aligned, the effective cone must be non-simplicial, which means that there will be additional instanton contributions to the axion potential.}
\label{nonsimplicial}
\end{center}
\end{figure}

Before we get to these constraints, however, we briefly comment on decay constant alignment for compactifications on Calabi-Yau's with $h^{1,1} \geq 3$.  In these cases, alignment can occur between irreducible effective divisors which each correspond to generators of the effective cone.  However, the requirement that the the effective cone contain the nef cone implies that whenever two generators experience near-perfect alignment, the effective cone must be non-simplicial, as shown in Figure \ref{nonsimplicial}.  A non-simplicial cone implies additional irreducible divisors wrapped by the Euclidean $D3$-branes, which means additional contributions to the potential which can cut off the range available to the inflaton.  Once again, contemporary knowledge of the effective cone precludes a more rigorous statement, but if the simulated examples of \cite{Rudelius} are any indication, it is not unreasonable to think that these additional contributions should become significant whenever the alignment is substantial.

\section{Alignment and the Weak Gravity Conjecture \label{sec:WGC}}

Similar to the cases of i) natural inflation \cite{Freese}, ii) extranatural inflation \cite{extranatural}, and iii) $N$-flation \cite{Nflation}, inflation by decay constant alignment seems to be resisting an embedding in string theory.  In \cite{WGC}, it was pointed out that the difficulty of embedding i) and ii) could be explained by the `generalized weak gravity conjecture.'  In \cite{Rudelius}, case iii) was shown to follow from the generalized $N$-species extension of the weak gravity conjecture of \cite{Cheung}.  In what follows, we show that this same extension applied to the case of two axion species places significant constraints on models of decay constant alignment.

To begin, we recall the statements of the relevant versions of the weak gravity conjecture\footnote{The inflationary implications of a related conjecture, also named the `weak gravity conjecture,' were initiated in \cite{Huang}, though that version of the conjecture is quite different from the ones studied here.}.  The weak gravity conjecture in its original and most basic form holds that in $4$ dimensions, any $U(1)$ gauge theory (or theory that can be Higgsed to a $U(1)$) that admits an ultraviolet embedding into a consistent theory of quantum gravity must have a particle of mass to charge ratio $M/q \leq 1$.  The reasoning behind this is both simple and compelling: if such a particle does not exist, there would be nothing to carry away charge from an extremal black hole without introducing a naked singularity, and so in avoiding this fate one would necessarily be left with the extremal black hole remnants, which introduce problems of their own \cite{Susskind}.

In \cite{WGC}, this original conjecture was generalized to $p$-form gauge fields in $d$-dimensions, where the natural conjecture is that there should be a $p-1$-dimensional object with tension $T \lesssim (g^2/G_N)^{1/2}$.  In particular, a $0$-form axion field with decay constant $f$ should give rise to a $-1$-dimensional object (instanton) with action $S \lesssim M_p/f$.  But then, considering the axion potential
\begin{equation}
V \sim \sum_n e^{-n S} \cos{n\phi/f},
\label{potentialform}
\end{equation}
we see that whenever $f$ becomes of order $N\,M_p$, terms up to order $n=N$ in the potential become non-negligible, and as a result one cannot achieve an inflaton traversal much larger than $M_p$.  This `generalized' version of the weak gravity conjecture does not have quite the same support as the original version coming from arguments regarding black hole remnants, but it would explain the difficulty observed in \cite{Banks} of attaining a parametrically large axion decay constant in string theory.

In \cite{Cheung}, it was pointed out that the $N$-species extension of the weak gravity conjecture is stronger than the statement that there must exist one particle with mass to charge ratio $M/q_i \leq 1$ for each $U(1)_i$ gauge symmetry.  Rather, the correct extension to $N$ $U(1)$'s is that there must exist a collection of particle species $i=1,...,N$ with charge vectors $\vec{q}_i$ and charge-to-mass vectors $\vec{z}_i = \vec{q}_i \frac{M_p}{m_i}$, and the convex hull spanned by the vectors $\pm \vec{z}_i$ must contain the $N$-dimensional unit ball.  Otherwise, one would again be left with unwanted black hole remnants.

The generalization of the $N$-species extension of the weak gravity conjecture to $0$-form axion fields is then natural: for a 0-form axion, the appropriate analog of the charge-to-mass vector is $\vec{z}_i = \sum_j \frac{M_p}{f_{ij} S_i} \vec{e}_j$, where the $\{\vec{e}_j\}$ form an orthonormal basis of the vector space.  The potential takes the form,
\begin{equation}
V \sim \sum_i \mathcal{A}_i e^{-S_i} \cos( \sum_j\frac{\phi_j}{f_{ij}}),
\label{multiV}
\end{equation}
with canonical kinetic terms for the fields $\phi_i$.  The generalized weak gravity conjecture holds that the convex hull spanned by these vectors and their negatives should contain the unit ball.  When the charge vectors are set orthogonal to each other, this conjecture says precisely that the radius of axion moduli space is $\lesssim M_p$ \cite{Rudelius}, ruling out the possibility of $N$-flation.

Here, we point out that this weak gravity conjecture also constrains parametric enhancement of the moduli space radius by means of decay constant alignment, and up to some reasonable assumptions it rules out enhancement altogether in the special case in which there are as many cosine terms in the potential (\ref{multiV}) as there are axions $\phi_i$.  Heuristically, as one starts to align the axion decay constants, the charge vectors $\vec{Z}_i$ become aligned.  But in order for the convex hull of the vectors $\pm \vec{z}_i$ to continue to span the unit ball, the vectors must get longer as the angle between them gets smaller.  In the end, these effects cancel each other out, so that the prospects of attaining a trans-Planckian inflaton traversal are not enhanced by decay constant alignment.

\begin{figure}
\begin{center}
\includegraphics[trim=7mm 7mm 7mm 7mm, clip, width=70mm]{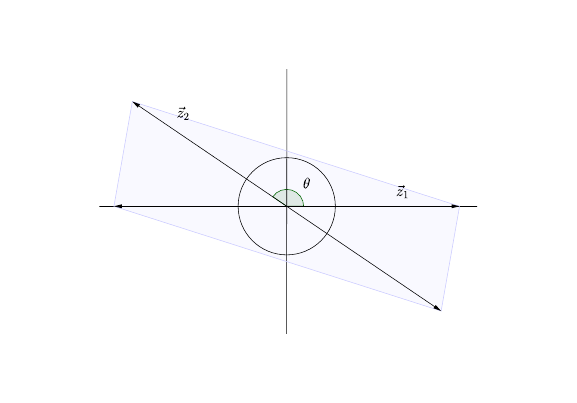}
\includegraphics[trim=7mm 7mm 7mm 7mm, clip, width=70mm]{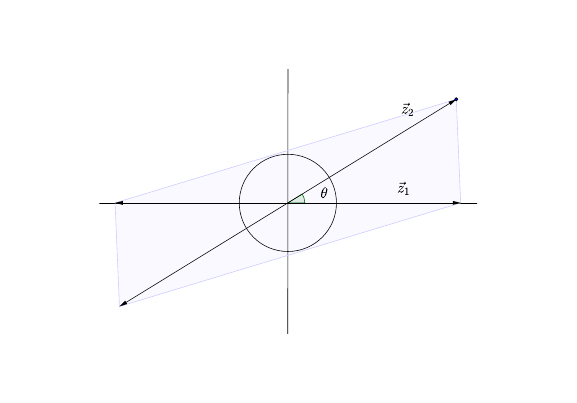}
\caption{The angle between the charge vectors approaches $\pi$ for anti-alignment (left) and $0$ for alignment (right).  The requirement that the convex hull spanned by the vectors and their negatives should contain the unit ball restricts the maximum displacement of the inflaton to be $\lesssim \pi M_p$.}
\label{chargevectorfigure}
\end{center}
\end{figure}

To be more precise, we begin with the necessary condition on the charge vectors $\vec{z}_1$, $\vec{z}_2$, so that the convex hull spanned by $\pm \vec{z}_i$ contains the 2-dimensional unit ball \cite{Cheung},
\begin{equation}
(|\vec{z}_1|^2-1)(|\vec{z}_2|^2-1) \geq (1+|\vec{z}_1 \cdot \vec{z}_2|)^2.
\label{hullconstraint}
\end{equation}
From our previous discussion, we expect that for an instanton action $S_i \sim N$, harmonics up to order $N$ will be appreciable.  Even if additional factors of $N$ suppress the higher harmonics (as is the case in extranatural inflation \cite{extranatural}), we still expect $S_i >1$ in order to suppress the potential relative to the Planck scale and yield an axion light enough for inflation.  This could also be avoided if $\mathcal{A}_i$ is much smaller than $M_p^4$, so the axion potential is significantly suppressed by some prefactor other than the instanton action $e^{-S_i}$.  However, one must also ensure that theoretical control of non-perturbative effects is maintained in this limit--a non-trivial challenge for string compactifications.  With these assumptions, we can simplify the analysis by setting $S_i \gtrsim 1$.  Further, we can without loss of generality rotate the charge vectors so that the first lies along the positive $x$-axis and the second either lies in the first quadrant (in the case of alignment) or in the second quadrant (in the case of anti-alignment).  With this, the charge vectors become
\begin{equation}
\vec{z}_1 = (\frac{M_p}{f_1},0)\,,~~~~\vec{z}_2 =( \frac{M_p}{f_2}\cos\theta,\frac{M_p}{f_2}\sin\theta),
\end{equation}
where $\theta$ is the angle between the two charge vectors, as shown in Figure \ref{chargevectorfigure}.  Decay constant alignment occurs when $\theta \rightarrow 0$, and anti-alignment occurs when $\theta \rightarrow \pi$.

The radius of moduli space, or equivalently the maximal displacement of the inflaton during slow-roll, is given by,
\begin{equation}
r \approx \left((f_1 \pi)^2 + (\pi/\sin\theta)^2(f_2 \pm f_1 \cos\theta)^2\right)^{1/2},
\end{equation}
with the $+$ sign corresponding to alignment and the $-$ sign corresponding to anti-alignment.  Further, the constraint (\ref{hullconstraint}) becomes,
\begin{equation}
(\frac{1}{f_1^2} -  1)(\frac{1}{f_2^2}-1) \gtrsim (1\pm\frac{1}{(f_1 f_2\cos\theta^2)}),
\end{equation}
where again the $+$ corresponds to alignment and the $-$ corresponds to anti-alignment.  We therefore have a problem of constrained maximization, which can be solved with the method of Lagrangian multipliers.  Solving, we find in both cases\footnote{This result was double-checked using numerical maximization.},
\begin{equation}
r \lesssim \pi M_p.
\label{rbound}
\end{equation}
In other words, the weak gravity constraint is, once again, precisely the statement that the moduli space radius is bounded by (roughly) $\pi M_p$.  As angle between the charge vectors goes to $0$ or $\pi$, the decay constants must shrink so that the unit ball remains in the span of the charge vectors and their negatives.  These effects exactly cancel each other out, in the sense that for any angle you choose, it is possible to find $f$ and $g$ such that the bound (\ref{rbound}) is saturated, but it is not possible to choose $f$ and $g$ such that it is violated.  The generalized weak gravity conjecture therefore forbids enhancement of the moduli space radius relative to $M_p$.  If the conjecture is correct, this would explain the difficulty observed in \cite{newMcAllister} of achieving an axion moduli space radius larger than $\pi M_p$, even in a compactification with $h^{1,1}=51$ where $N$-flation would n\"aively have predicted a radius of size $\sqrt{51}\pi M_p$ and decay constant alignment could have allowed even greater enhancement.  Note that this does not contradict the claims of \cite{newMcAllister} that the axion moduli space radius can scale by as much as $(h^{1,1})^{3/2}$ relative to a typical axion decay constant $f$ in the compactification.  In such a situation, $f$ would scale as $(h^{1,1})^{-3/2} M_p$, and the two factors would cancel each other out so that the radius stays bounded by $\pi M_p$.  This is precisely what has occurred in the aforementioned example with $h^{1,1}=51$.

\begin{figure}
\begin{center}
\includegraphics[trim=13mm 10mm 15mm 8mm, clip, width=80mm]{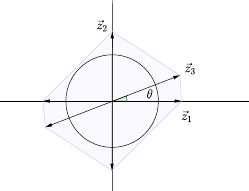}
\caption{A model with three charge vectors and two axions.  Although the generalized weak gravity conjecture still constrains the size of moduli space, one could conceivably achieve a large inflaton traversal by taking $\theta \rightarrow 0$ as long as the potential contributions from $\vec{z}_3$ dominate those from $\vec{z}_2$, thereby allowing multiple traversals of the $\phi_2$ axion moduli space.}
\label{3vector}
\end{center}
\end{figure}

However, although the generalized weak gravity conjecture holds that moduli space radius is necessarily $\lesssim \pi M_p$, it does not rule out the possibility that additional potential contributions could dominate over the ones henceforth considered and allow for the axions to traverse their moduli space multiple times, leading to prolonged inflation.  In fact, it is even possible that decay constant alignment could play a role in facilitating these traversals\footnote{We are grateful to P. Saraswat and R. Sundrum for pointing this out to us.  See \cite{Sundrum} for a realization of decay constant alignment obeying the weak gravity conjecture in the context of extranatural inflation.}.  Consider a toy model (shown in Figure \ref{3vector}) with two axions but three cosine terms in the potential (\ref{multiV}), whose charge vectors are given by
\begin{equation}
\vec{z}_1=(z_1,0)\,,~~~~\vec{z}_2=(0,z_2)\,,~~~~\vec{z}_3=z_3(\cos{\theta}, \sin{\theta}).
\end{equation}
In this basis, the decay constant matrix simplifies to
\begin{equation}
f_{ij} = \left( \begin{array}{cc}
f_1 & 0  \\
0 & f_2 \\
f_{31} & f_{32} \end{array} \right).
\end{equation}
Once again, the generalized weak gravity conjecture holds that the convex hull spanned by these vectors and their negatives should contain the unit ball.  When there were only two vectors, we found strong constraints on the radius of axion moduli space, ruling out the possibility of parametrically-prolonged slow-roll inflation.  Indeed, the moduli space radius is once again limited here, but nonetheless prolonged inflation may still be possible.  If we set $z_1 = z_2 = \sqrt{2}$, then we have already satisfied the constraints of the weak gravity conjecture and restricted the size of the axion moduli space to be $\mathcal{O}(\pi M_p)$, but we have left $\vec{z}_3$ unconstrained.  If we take $\theta \rightarrow 0$ and find a way to ensure that the potential contributions associated with $\vec{z}_1$ and $\vec{z}_3$ dominate over those from $\vec{z}_2$, then decay constant alignment will occur between the dominant potential contributions, and a parametrically-large inflaton traversal can be realized.

However, it is non-trivial to ensure that the potential contributions associated with $\vec{z}_1$ and $\vec{z}_3$ dominate over those from $\vec{z}_2$ in a realistic axion inflation scenario.  To match with experiment, the inflaton field cannot be heavier than about $10^{-5} M_p$.  In our toy model with $\mathcal{A}_i \sim M_p^4$ and $\theta \rightarrow 0$, this would mean setting the action $S_3 \overset{\sim}{\rightarrow} 20$, so that the potential contribution $\mathcal{A}_3 e^{- S_3} \cos(\sum_j \frac{\phi_j}{f_{3j}}) \subset V$ yields a sufficiently light axion $\phi_2$.  But then, in order for this potential contribution to dominate over the one from $\vec{z}_2$, we must have $S_2 \gtrsim 20$.  Since we also needed $z_2 \gtrsim \sqrt{2}$, we must have $f_2 \lesssim \frac{1}{20\sqrt{2}} M_p$.  Further, the usual constraint $S_1 \gtrsim 1$ implies $f_1 \lesssim \frac{1}{\sqrt{2}} M_p$.

Now, we specialize to the case of $\vartheta_i$ axions in type IIB compactifications considered earlier (though this analysis applies rather generally to axions in string theory).  Here, the potential contributions under inspection come from Euclidean $D3$-branes wrapping irreducible 4-cycles.  Therefore, we can expand the divisor $\tilde{D}_3$ corresponding to $\vec{z}_3$ in the basis $\{\tilde{D}_1, \tilde{D}_2\}$ corresponding to $\vec{z}_1, \vec{z}_2$, respectively.  This yields additional relations between the actions $S_i$ (which are given classically up to a factor of $g_s$ by the volumes of the $\tilde{D}_i$) and the decay constants $f_{ij}$,
\begin{equation}
S_3 = a S_1 + b S_2\,,~~~~(\frac{1}{f_{31}},\frac{1}{f_{32}}) = (\frac{a}{f_1},\frac{b}{f_2}). 
\label{stringrelation}
\end{equation}
The relevant piece of information from string theory is that the same coefficients $a,b$ appear in both of the equations in (\ref{stringrelation}), at least at the classical level.

We now want to achieve parametric enhancement of the field range accessible to the $\phi_2$ axion during inflation by taking $\theta \rightarrow 0$.  The farthest point accessible to this axion is found by simultaneously solving the equations,
\begin{equation}
\frac{1}{f_1}\phi_1 = \pi\,,~~~~  \frac{a}{f_1}\phi_1 +\frac{b}{f_2}\phi_2 = \pi.
\end{equation}
This gives $|\phi_2| = \frac{\pi f_2 a}{b}$.  But now, enforcing $f_2 \lesssim \frac{1}{20\sqrt{2}} M_p$ and requiring the field range to be $\gtrsim 15 \, M_p$ to satisfy observational constraints implies $\frac{b}{a} \lesssim \frac{1}{100}$.  Recalling that $\tilde{D}_3 = a \tilde{D}_1 + b \tilde{D}_2$, we conclude that the level of alignment needed for inflation in our toy model would require an impressive hierarchy between the coefficients $a$ and $b$.  As discussed in \S \ref{sec:STRING}, such a hierarchy is very difficult to achieve while maintaining integrality of the triple intersection numbers, and it cannot be expected to occur in a typical compactification, if ever.  A more promising method for realizing axion inflation would be to tune the prefactors $\mathcal{A}_i$ so that $\mathcal{A}_1, \mathcal{A}_3 \ll M_p^4$, $\mathcal{A}_2 < \mathcal{A}_1, \mathcal{A}_3$.  In this case, we would not need $S_2$ to be as large, and the ratio $\frac{b}{a}$ can be much more modest.

The conclusion of this calculation agrees well with our study of $\vartheta_i$ axion decay constant alignment in type IIB Calabi-Yau compactifications with $h^{1,1}=2$.  There, we saw that (neglecting the irreducible divisors in the interior of the effective cone), one can not achieve inflation through decay constant anti-alignment.  If there are only two irreducible effective divisors, the radius of moduli space is fixed to be $r < \pi \sqrt{\frac{3}{2}} M_p$.  Similarly here, if there are only two charge vectors (each of which corresponds to an irreducible effective divisor), then the generalized weak gravity conjecture forbids decay constant alignment, and bounds the radius of moduli space at $r \lesssim \pi M_p$.  If, however, one allows for additional irreducible effective divisors, one can conceivably still realize decay constant alignment.  Similarly here, once additional charge vectors are included, the generalized weak gravity conjecture no longer rules out axion inflation.  But in order to get a field range suitable for inflation, the degree of alignment must be rather strong ($\theta \ll 1$), and the hierarchy of scales must be balanced so that the potential contributions from the generating divisors of the effective cone/charge vectors which satisfy the weak gravity conjecture do not dominate the potential.  Accomplishing both of these feats is not easy, so even if the weak gravity conjecture does not rule out decay constant alignment in string theory, it nonetheless imposes significant constraints on the compactification geometry.

The preceding arguments relied upon the $0$-form generalization of the weak gravity conjecture, which is not based on arguments regarding black holes remnants and hence is more conjectural.  However, it is possible to relate this generalized weak gravity conjecture to the original version regarding black hole remnants for axions descending from $p$-form gauge fields wrapped on $p$-cycles, as is the case in extranatural inflation and most stringy models of axion inflation (in particular the type IIB models involving axions descending from NS-NS or R-R $p$-forms considered here) \cite{ReeceRudelius}.

\section{Decay Constant Alignment from $D7$-Branes \label{sec:D7}}

In \cite{McAllisterLong}, an alternate scenario for decay constant alignment of $\vartheta_i$ axions in type IIB compactifications using spacetime-filling $D7$-branes rather than Euclidean $D3$-brane instantons was presented.  In this section, we point out that this scenario a) could avoid the constraints of the current paper and b) could also even avoid the original bound of \cite{Banks} on axion decay constants in string theory.  However, we show that there are other issues facing this model of decay constant alignment, so the challenge of realizing natural inflation with $\vartheta_i$ axions in type IIB compactifications remains an open problem.

Gaugino condensation on $N$ $D7$-branes wrapping $m$ times around a $4$-cycle of a Calabi-Yau $Z$ produces potential contributions of the form,
\begin{equation}
V \supset e^{-\frac{m}{N} \tau} \cos(\frac{m}{N} \vartheta).
\label{D7potential}
\end{equation}
The crucial ingredient which allows for parametric enhancement of the inflaton traversal here is the fact that $D7$-brane state is a classical configuration and does not yield a tower of instantons.  Thus, we are allowed to specify a single winding number $m$ rather than summing over all positive integers $m$.  From our previous discussion, it is not difficult to see how this could lead to decay constant alignment: the primary challenge to achieving alignment in \S 3 was the additional contributions to the potential that came from either Euclidean $D3$-branes wrapping the wrong irreducible cycles or else wrapping the correct cycles the wrong number of times.  With $D7$-branes, we are not required to sum over all possible ways to wrap the cycles, so we can tune the wrappings in such a way so as to avoid these difficulties.  Furthermore, the added degree of freedom $N$, which counts the number of branes wrapping the cycle, allows us to tune the masses of the axions to make them suitable inflaton candidates.  This would allow decay constant alignment to be realized on just about any compactification geometry at the expense of only a discrete tuning of the winding number of the $D7$-branes.  An explicit example of decay constant alignment was worked out in \cite{McAllisterLong}.

This scenario also avoids the arguments of \S 4.  There, the no-go result derived from the fact that (according to the generalized weak gravity conjecture) the instanton coupled to the axion must become light whenever the axion decay constant grows larger than $M_p$.  Here, whenever there are $D7$-branes wrapping a cycle, there will also be Euclidean $D3$-branes wrapping the cycle, so that indeed the shift symmetry will be broken over a range $\lesssim M_p$ by these instanton effects.  However, if the leading contribution to the axion potential comes from the $D7$-branes rather than the Euclidean $D3$-brane instantons, then the instanton effects prescribed by the weak gravity conjecture will be sub-dominant.  As a result, the inflaton will traverse its moduli space multiple times as it relaxes to the minimum of its potential.  Similar to axion monodromy inflation \cite{AxionMonodromy}, this could allow for a trans-Planckian inflaton traversal without violating the bounds of the weak gravity conjecture.

It is worth noting that this scenario does not require two axions to generate a large inflaton traversal--it can be used to generate parametric enhancement of a single axion decay constant as well.  One simply takes sets the winding number $m=1$ and takes the number of branes $N$ wrapping the cycle to be large.  Then, the axion decay constant of the axion in (\ref{D7potential}) becomes,
\begin{equation}
f = g N,
\end{equation}
where $-\frac{1}{2}g^2 \partial \vartheta^2$ is the metric on $\vartheta$ moduli space.  Once we have fixed $\tau$ large enough so that the dominant contributions to the potential come from $D7$-branes rather than Euclidean $D3$-branes, $g$ is completely determined, so we realize parametric enhancement in the axion decay constant by taking $N$ large.  Note that the mass of the axion goes as,
\begin{equation}
m^2 \approx e^{- \tau/N}/(g N)^2,
\end{equation}
so that we can even attain parametric suppression of the mass at fixed $\tau, g$ by taking $N$ large.

However, a different geometric issue faces this model of decay constant alignment.  As pointed out in \cite{McAllisterLong}, intersecting $D7$-branes will give rise to light $W$-bosons which must be made massive in order for the necessary potential contributions in (\ref{D7potential}) to be present.  Thus, the $D7$-branes wrapping two effective cycles $\hat{D}_1$, $\hat{D}_2$ must not intersect.  Two curves wrapping a torus with identical winding numbers provide an example of this: such curves can be positioned so that they do not intersect each other.

However, curves on a torus can be misleading when it comes to intersection theory, because curves on a torus have a skew-symmetric intersection form rather than a symmetric one\footnote{We are grateful to D. Morrison and M. Esole for discussions on this point.}.  If the two divisors $\hat{D}_1$ and $\hat{D}_2$ do not intersect, then their intersection class $\hat{D}_1 \cdot \hat{D}_2$ must vanish, so that $\hat{D}_1 \cdot \hat{D}_2 \cdot D =0$ for all divisors $D$.  Expanding in a basis of divisors $\{ D_1, D_2\}$ by setting $\hat{D}_1 = a_{11} D_1 + a_{12} D_2$, $\hat{D}_2 = a_{21} D_1 + a_{22} D_2$, the requirement that the two $D7$-branes do not intersect gives rise to the condition,
\begin{equation}
(a_{11} D_1 + a_{12} D_2) \cdot (a_{21} D_1 + a_{22} D_2) \cdot D_i = 0\,,~~~~i=1,2.
\label{inteq1}
\end{equation}
However, in order for the $D7$-brane potential contributions to the $\vartheta_i$ axions to dominate over the Euclidean $D3$-brane potential contributions wrapping the same cycles, there must be multiple $D7$-branes wrapping each cycle.  Requiring that these $D7$-branes do not intersect imposes the conditions,
$$
(a_{11} D_1 +a_{12} D_2) \cdot (a_{11} D_1 + a_{12} D_2) \cdot D_i = 0\,,~~~~i=1,2,
$$
\begin{equation}
(a_{21} D_1 + a_{22} D_2) \cdot (a_{21} D_1 + a_{22} D_2) \cdot D_i = 0\,,~~~~i=1,2.
\label{inteq2}
\end{equation}
Note that the intuition provided by curves wrapping a torus would have led us astray here--the fact that the intersection form on a torus is skew-symmetric means that two curves wrapping the same cycles on a torus will have zero intersection class and so intersections can be avoided.  This is not the case for 4-cycles on a 6-dimensional Calabi-Yau.  Solving (\ref{inteq1}) and (\ref{inteq2}) simultaneously yields,
\begin{equation}
\frac{a_{12}}{a_{11}} = \frac{a_{22}}{a_{21}} \mbox{ OR } \kappa_{111}=\kappa_{112}=\kappa_{122}=\kappa_{222}=0.
\label{solns}
\end{equation}
$\frac{a_{12}}{a_{11}} = \frac{a_{22}}{a_{21}} $ would imply perfect alignment, which leaves some of the axionic shift symmetry unbroken and is therefore insufficient.  $\kappa_{111}=$ $\kappa_{112}=$ $\kappa_{122}=$ $\kappa_{222}=0$ is clearly unacceptable.  It is therefore not possible to achieve decay constant alignment here using $D7$-branes which do not intersect.

Nonetheless, it may be possible to achieve this model of alignment using Calabi-Yau manifolds with greater Hodge number $h^{1,1}$.  The generalization of (\ref{solns}) to these scenarios is,
\begin{equation}
\kappa_{11i}=\kappa_{12i}=\kappa_{22i}=0\,,~~~~i=1,...,h_+^{1,1}.
\end{equation}  
This is a rather strong constraint--it implies that there must exist a plane in $N^{1}(Z)$ spanned by two effective divisors such that the intersection of any two divisors on this plane is the zero homology class.  We do not expect this condition to hold for a generic Calabi-Yau, if ever, but we cannot immediately rule it out as we can with $h^{1,1}=2$ (or $h^{1,1}=3$, where the metric necessarily becomes singular once this constraint is imposed).  Further, dominance of the $D7$-brane contributions over the Euclidean $D3$-brane contributions requires large ranks for the condensing gauge groups, and it has been argued \cite{Conlon} that this will lead to large $N$ species problems such as violations of entropy bounds.  Nonetheless, in light of the bound hypothesized by the generalized weak gravity conjecture, this scenario is certainly worth considering further.  If one could find an example of a Calabi-Yau satisfying the prerequisite conditions, one could overcome the constraints of the weak gravity conjecture and achieve inflation in a relatively straightforward manner.  Most assuredly, the underlying philosophy behind this scenario--introducing additional potential contributions which dominate over the ones from Euclidean $D3$-branes satisfying the weak gravity conjecture--is the most promising route for achieving $\vartheta_i$ axion inflation in string theory.

\section{Final Remarks}

The arguments presented in this paper do not represent a proof against the possibility of natural axion inflation by means of decay constant alignment.  We have seen that in type IIB string theory, it is conceivable that one could get around the no-go theorem facing the anti-alignment mechanism by an alignment mechanism in which additional potential contributions are introduced and the inflaton traverses its moduli space multiple times, though such a phenomenon seems unlikely to occur in a typical compactification.  The generalized weak gravity conjecture similarly forbids parametric enhancement of the moduli space radius through simple alignment scenarios, but it does not rule out the aforementioned type IIB alignment mechanism involving additional potential contributions and multiple windings of the inflaton.  Unlike the original weak gravity conjecture, the generalized version does not derive from arguments regarding extremal black hole remnants, so it is relatively more conjectural at present.  However, there is recent progress in this direction \cite{ReeceRudelius, Shiu}.

These arguments deal exclusively with axionic shift symmetries which are broken by instanton effects.  The scenario of \cite{McAllisterLong} admits a periodic potential for the axions from $D7$-branes wrapping appropriate effective divisors in the Calabi-Yau manifold, and so avoids some of the constraints.  However, there are other issues with this model, which can only be alleviated at the expense of imposing strong geometric constraints on the Calabi-Yau.  Satisfying these constraints is manifestly impossible in Calabi-Yau's with $h^{1,1} \leq 3$ and seems rather unlikely in Calabi-Yau's with greater $h^{1,1}$.

There is another possible loophole in our arguments which could be interesting to explore more fully in the context of quantum gravity.  Suppose the axion potential takes the form:
\begin{equation}
V = \sum_n \mathcal{A} \frac{1}{n^\alpha} e^{-n a S} \cos{n\phi/f} ,
\end{equation}
So far, we have usually assumed $a \sim 1$, $\alpha=0$, and $\mathcal{A} \sim M_p^4$. If $a \gg 1$, then one could achieve an axion decay constant of order $a \, M_p$ without introducing higher order harmonics to the potential or violating the generalized weak gravity conjecture.  However, this parametric enhancement does not arise among the common axions of string theory or in QCD, and so it seems unlikely.  On the other hand, if $\alpha > 1$ and $\mathcal{A} \sim 10^{-10} M_p^4$, then even if $a \sim 1$, one could still achieve successful inflation and satisfy the constraints of the weak gravity conjecture by taking $S \ll 1$ and using an alignment mechanism to produce an effective $f \gg M_p$.  This works because the higher harmonics will still be suppressed by factors of $n^{-\alpha}$ and the inflaton mass will still be suppressed by the factor of $\mathcal{A}$.  Indeed, it is possible to arrange for such a limit in extranatural inflation by taking the compactification radius large\footnote{We are grateful to M. Reece, P. Saraswat, and R. Sundrum for pointing this out to us.}, but in the case of type IIB $\vartheta_i$ axions the compactification volume cannot be taken large unless the action $S \sim Vol(D)$ is taken large as well, disfavoring this possibility.  Furthermore, the limit $S \ll 1$ would imply $Vol(D) \ll 1$ (in string units), taking us out of the large volume limit and into the realm where metric corrections are not under theoretical control.  Nonetheless, it would be interesting to see if these loopholes can be somehow exploited in string theory.

If the generalized weak gravity conjecture is correct and such loopholes do not arise, the phenomenological implications for cosmology are rather interesting.  If axions are responsible for inflation, the breaking of their shift symmetry at leading order must be accomplished by either instantons immune to the weak gravity conjecture (as discussed in \S \ref{sec:WGC}) or by other effects such as fluxes (as in axion monodromy inflation) or $D$-branes (as in the model discussed in \S \ref{sec:D7}).  However, subleading instanton effects are bound to arise (in order to satisfy the weak gravity conjecture) and modulate the potential \cite{AxionMonodromy2}, thereby leading to resonant non-Gaussianity \cite{resonant} which may be detected in upcoming experiments of the CMB or LSS.  Furthermore, the magnitude of resonant non-Gaussianity is affected by the frequency of the modulations, so bounds on axion decay constants translate to constraints on non-Gaussianity.

String theory appears to be resisting an embedding of natural inflation, even if one entertains the possibility of using multiple axions and/or attempting to align their decay constants.  And, remarkably though perhaps unsurprisingly, the generalized weak gravity conjecture agrees with string theory in these instances.  The evidence for this generalization of the weak gravity conjecture is therefore growing quickly, and on top of the evidence provided in the original paper \cite{WGC} and more recently in \cite{WGCin3d}, the arguments here regarding the $0$-form axion version of the conjecture give us good reason to think that the weak gravity conjecture has not unveiled all of its secrets yet.  Further studies of this mysterious conjecture are clearly in order.

From an even broader perspective, the arguments discussed here indicate that the effects of quantum gravity reach further than one might have expected.  While the list of inflationary field theories consistent with observation is hopelessly large (see e.g. \cite{encyclopedia}), the list of inflationary field theories that are also consistent with ultraviolet physics may be more manageable.  The surprising tension between the various models of axion inflation (which are in principle supposed to be more compatible with high-energy physics than most theories of inflation due to their shift symmetries) and string theory suggests that the `swampland' \cite{swampland} of inflationary theories might be larger than previously expected.
\\
\\
We wish to thank T. Bachlechner, M. Esole, C. Long, L. McAllister, D. Morrison, M. Reece, P. Saraswat, R. Sundrum, and C. Vafa for helpful discussions.  We also thank the 2015 Physics and Geometry of F-Theory Workshop at the Max Planck Institute in Munich for hospitality during parts of this work.  This material is based upon work supported by the National Science Foundation under Grant No. DGE-1144152.


\begin{thebibliography}{9} 

\bibitem{guth}
A.H. Guth, Phys.Rev. D23, 347 (1981).

\bibitem{linde}
A.D. Linde, Phys.Lett. B108, 389 (1982).

\bibitem{albrecht&steinhardt}
A. Albrecht and P.J. Steinhardt, Phys.Rev.Lett. 48, 1220 (1982).

\bibitem{bicep2}
P. Ade \emph{et al.} (BICEP2 Collaboration), Phys.Rev.Lett. 112, 241101 (2014), arXiv:1403.3985 [astro-ph.CO].

\bibitem{axion1}
M. Czerny, T. Higaki, and F. Takahashi, JHEP 1405 144 (2014), arXiv:1403.0410 [hep-ph].

\bibitem{axion2}
M. Czerny, T. Higaki, and F. Takahashi, Phys. Lett. B 167-172 (2014), arXiv:1403.5883 [hep-ph].

\bibitem{axion3}
K. Choi, H. Kim, and S. Yun, Phys. Rev. D 90 023545 (2014), arXiv:1404.6209 [hep-th].

\bibitem{axion4}
S.-H. H. Tye and S. S. C. Wong, (2014), arXiv:1404.6988 [astro-ph.CO].

\bibitem{axion5}
R. Kappl, S. Krippendorf, and H. P. Nilles, (2014), arXiv:1404.7127 [hep-th].

\bibitem{axion6}
I. Ben-Dayan, F. G. Pedro, and A. Westphal,  (2014), arXiv:1404.7773 [hep-th].

\bibitem{McAllisterLong}
C. Long, L. McAllister, P. McGuirk, Phys. Rev. D 90, 023501 (2014), arXiv:1404.7852 [hep-th].

\bibitem{axion8}
T. Li, Z. Li, and D. V. Nanopoulos, (2014), arXiv:1407.1819 [hep-th].

\bibitem{Pramod}
X. Gao, T. Li, P. Shukla, JCAP 1410, 048 (2014), arXiv:1406.0341 [hep-th].

\bibitem{axion9}
Z. Kenton and S. Thomas, (2014), arXiv:1409.1221 [hep-th].

\bibitem{axion10}
T. Ali, S. S. Haque and V. Jejjala, (2014), arXiv:1410.4660 [hep-th].

\bibitem{axion11}
T. Higaki and F. Takahashi, (2014), arXiv:1409.8409 [hep-th].

\bibitem{axion12}
H. Abe, T. Kobayashi, and H. Otsuka, (2014), arXiv:1411.4768 [hep-th].

\bibitem{axion13}
Y. Bai and B. A. Stefanek, (2014), arXiv:1405.6720 [hep-ph].

\bibitem{Sundrum}
A. de la Fuente, P. Saraswat, and R. Sundrum, (2014), arXiv:1412.3457 [hep-th].

\bibitem{Planckdust}
 R. Adam \emph{et al.} (\emph{Planck} Collaboration), (2014), arXiv:1409.5738 [astro-ph.CO].

\bibitem{PlanckBicep}
P. Ade \emph{et al.} (BICEP2/Keck and \emph{Planck} Collaborations), (2015), arXiv:1502.00612 [astro-ph.CO].

\bibitem{Zaldarriaga}
P. Creminelli, D. L. Nacir, M. Simonović, G. Trevisan, and M. Zaldarriaga, (2015), arXiv:1502.01983 [astro-ph.CO].

\bibitem{Lotfi}
L. Boubekeur, (2013), arXiv:1312.4768 [astro-ph.CO].

\bibitem{Banks}
T. Banks, M. Dine, P.J. Fox, and E. Gorbatov, JCAP 0306, 001 (2003), arXiv:0303252 [hep-th].

\bibitem{Nflation}
S. Dimopoulos, S. Kachru, J. McGreevy, and J.G. Wacker, JCAP 0808, 003 (2008), arXiv:0507205 [hep-th].

\bibitem{AxionMonodromy}
L. McAllister, E. Silverstein, and A. Westphal, Phys. Rev. D 82 046003 (2010), arXiv:0808:0706 [hep-th].

\bibitem{KNP}
J.E. Kim, H.P. Nilles, M. Peloso, JCAP 0501, 005 (2005), arXiv:0409138 [hep-ph].

\bibitem{Rudelius}
T. Rudelius, (2014), arXiv:1409.5793 [hep-th].

\bibitem{CicoliDutta}
M. Cicoli, K. Dutta, and A. Maharana (2014), arXiv:1401.2579 [hep-th].

\bibitem{BaumannMcAllisterBook}
D. Baumann and L. McAllister, \emph{Inflation and String Theory}. Cambridge University Press: Cambridge (2014).

\bibitem{Halverson}
M. Cvetič, I. García-Etxebarria, and J. Halverson, Fortsch.Phys. 59, 243-283 (2011) arXiv:1009.5386 [hep-th].

\bibitem{Grimm}
T.W. Grimm, Phys. Rev. D77, 126007 (2008), arXiv:0710.3883 [hep-th].

\bibitem{KreuzerSkarke}
M. Kreuzer and H. Skarke, Adv. Theor. Math. Phys. 4 (2002), arXiv:0002240 [hep-th].

\bibitem{WittenMF}
E. Witten, Nucl.Phys.B471 195-216 (1996), arXiv:9603150 [hep-th].

\bibitem{MorrisoncitedinWitten}
P.S. Aspinwall, B.R. Greene, and D.R. Morrison, Nucl.Phys.B420, 184-242 (1994), arXiv:9311042 [hep-th].

\bibitem{Freese}
K. Freese, J.A. Frieman, and A.V. Olinto, Phys. Rev. Lett. 65, 3233 (1990).

\bibitem{extranatural}
N. Arkani-Hamed, H.C. Cheng, P. Creminelli, and L. Randall, Phys. Rev. Lett. 90, 221302 (2003), arXiv:0301218/hep-th.

\bibitem{WGC}
N. Arkani-Hamed, L. Motl, A. Nicolis, and C. Vafa, JHEP 0706, 060 (2007), arXiv:0601001 [hep-th].

\bibitem{Cheung}
C. Cheung and G.N. Remmen, Phys. Rev. Lett. 113, 051601 (2014), arXiv:1402.2287 [hep-ph].

\bibitem{Huang} 
  Q.G. Huang, JHEP 0705, 096 (2007),  arXiv:0703071 [hep-th].

\bibitem{Susskind}
L. Susskind, SU-ITP-95-1 (1995), arXiv:9501106 [hep-th].

\bibitem{newMcAllister}
T. Bachlechner, C. Long, L. McAllister, (2014), arXiv:1412.1093 [hep-th].

\bibitem{ReeceRudelius}
B. Heidenreich, M. Reece, and T. Rudelius, (2015), arXiv:1506.03447 [hep-th].

\bibitem{Conlon}
J. Conlon, JCAP 1209, 019 (2012), arXiv:1203.5476 [hep-th].

\bibitem{Shiu}
J. Brown, W. Cottrell, G. Shiu and P. Soler, (2015), arXiv:1503.04783 [hep-th].

\bibitem{AxionMonodromy2}
R. Flauger, L. McAllister, E. Pajer, A. Westphal, and G. Xu, JCAP 1006 009 (2010), arXiv:0907.2916 [hep-th].

\bibitem{resonant}
R. Flauger and E. Pajer, JCAP 1101 017 (2011), arXiv:1002.0833 [hep-th].

\bibitem{WGCin3d}
M. Li, W. Song, and T. Wang, JHEP 0603:094 (2006), arXiv:0601137 [hep-th].

\bibitem{encyclopedia}
J. Martin, C. Ringeval, and V. Vennin, (2006), arXiv:1303.3787 [astro-ph].

\bibitem{swampland}
C. Vafa, HUTP-05/A043 (2005), arXiv:0509212 [hep-th].

\end{thebibliography}
\end{document}